\documentclass[3p,times]{elsarticle}
\usepackage{ecrc}

\volume{00}

\firstpage{1}

\journalname{Science Bulletin}

\runauth{S. Longhi}

\usepackage{amssymb}
\usepackage[figuresright]{rotating}
\usepackage{color}

\begin{document}

\begin{frontmatter}



\title{Time reversal of a discrete system coupled to a continuum based on non-Hermitian flip}


\author{Stefano Longhi}
\address{Dipartimento di Fisica, Politecnico di Milano and Istituto di Fotonica e Nanotecnologie del Consiglio Nazionale delle Ricerche, Piazza L. da Vinci 32, I-20133 Milano, Italy\\ Tel/Fax: 0039 022399 6156/6126, email: longhi@fisi.polimi.it}

\begin{abstract}
Time reversal in quantum or classical systems described by an Hermitian Hamiltonian  is a physically allowed process, which requires in principle inverting the sign of
the Hamiltonian.  Here we consider the problem of time reversal of a  subsystem of discrete states coupled to an external environment characterized
by a continuum of states, into which they generally decay. It is shown that, by flipping the discrete-continuum coupling from an Hermitian to a non-Hermitian interaction, thus resulting in a non unitary dynamics, time reversal of the subsystem of discrete states can be achieved, while the continuum of states is not reversed. Exact time reversal requires frequency degeneracy of the discrete states, or large frequency mismatch among the discrete states as compared to the strength of indirect coupling mediated by the continuum. Interestingly, periodic and frequent switch of the discrete-continuum coupling results in a frozen dynamics of the subsystem of discrete states. 
\end{abstract}

\begin{keyword}
non-Hermitian dynamics; time reversal; Loschmidt echo
\end{keyword}

\end{frontmatter}

\section{Introduction}
The physics of open quantum and classical systems is 
of great relevance in different areas of science \cite{r1,r2,r3,r4,r5,r6,r7,r8,r9,r10,referee1,referee2,r11} ranging from atomic and molecular physics \cite{r5,r7,r8} to nuclear physics \cite{r1,r3,r4,r6}, dissipative systems driven out of equilibrium and quantum thermodynamics \cite{r12,r13}, quantum computing \cite{r7,r14,r15}, transport  in nanoscale and mesoscopic solid state devices \cite{r9,r10,r16,r17,r18,r19,r20,r21,r22}, optics and photonics \cite{r23,r24,r25,r26,r27}, and transport in biological structures \cite{r28,r29,r30}. More recently, non-Hermitian models have been also introduced  in {\em ab-initio} theories to provide complex extensions of quantum mechanics \cite{r31,r32,r33,r34,r35,r36,r37}. Studies of open quantum systems date back to some pioneering theoretical works, notably of Gamow,  Weisskopf-Wigner and Feshbach on
nuclear $\alpha$-decay \cite{r1},  spontaneous emission \cite{r2}, and nuclear reactions \cite{r3}, and are nowadays of continuous interest in a wide range of different areas of physics. In a typical situation, a discrete system
described by $N$ states is coupled to an external environment characterized
by a continuum of states, into which they generally decay. An important example if provided, for instance, by electronic or photonic transport in mesoscopic or macroscopic cavities with attached wave guides. A simple approach to describe open systems is to eliminate the continuum
degrees of freedom via Feshbach projection technique or similar methods \cite{r2,r3,r4,r5,r6,r8,r9}, leading to an effective non-Hermitian Hamiltonian description for the finite-dimensional $N$ subspace of discrete states. Such an approach turns out to be very effective in predicting important interference effects between discrete states and the continuum, such as Fano resonances, the existence of bound states in the continuum, external mixing of states, nonadiabatic dynamical transitions near exceptional points, nonlinear effects, etc. (see, for example, \cite{r10,referee1,referee2} and references therein for a comprehensive overview). A major question is whether one can time-reverse the dynamics in the {\em reduced} $N$-dimensional subspace of discrete states. For example, is it possible to reverse the decay process of an initial excitation of the $N$ discrete states into the continuum of states, so as to retrieve the initial excitation in the subspace of discrete states? The problem of time reversal in systems with many degrees of freedom is a longstanding one in physics and dates back to the famous controversy between Loschmidt and Boltzmann about the second law of thermodynamics and the arrow of time, i.e. how macroscopic irreversibility can arise even if dynamical equations of motion are time reversible. In principle, performing time reversal in an Hamiltonian system would `simply` require inverting the sign of
the Hamiltonian. Quantum or classical reversibility in presence of various perturbations has been actively studied in recent
years and is now described through the so-called Loschmidt echo \cite{r38,r39,r40}. In particular, time reversal dynamics  has been experimentally demonstrated in a wide variety of quantum and classical systems, including spin systems \cite{r41}, matter waves \cite{r42,r43}, acoustic \cite{r44}, electromagnetic \cite{r45} and water waves \cite{r46}. Reversing the sign of the Hamiltonian, however, is not a simple task, especially when infinite degrees of freedom like those of a continuum of states are involved and perturbations can rapidly deteriorate the Loschmidt echo.\\
In this work it is suggested that reversing the dynamics of a discrete system coupled to a continuum can be achieved by flipping the discrete-continuum interaction from an Hermitian to a non-Hermitian coupling. Physical models and practical implementations of non-Hermitian couplings have been considered in some recent works \cite{r47,r48,r49,r50,r51,r52,r52bis}, especially in the context of tight-binding lattice models. A non-Hermitian coupling does not generally conserve the probability of the discrete-continuum system, since another hidden reservoir is involved in the dynamics from which energy and/or particles can be fed or extracted. As far as time reversal dynamics is concerned, flipping the coupling of the discrete states with the continuum from an Hermitian to a non-Hermitian one can reverse, under certain conditions, the dynamics of the discrete states, but not the one of the continuum which can take energy from the hidden reservoir. Remarkably, periodic flipping of the coupling over a short time scale can result in a frozen dynamics of the $N$ discrete states, but not of the continuum of states. As an example, we discuss time reversal and frozen dynamics based on non-Hermitian flip in a multilevel Fano-Anderson model describing the coupling of $N$ sites to a tight-binding continuum. 

\section{Time reversal based on non-Hermitian flip}
\subsection{Basic model and effective Hamiltonian}
Let us consider a discrete system of $N$ states $|a_n \rangle$ with frequencies $\omega_n$ ($n=1,2,...,N$) which are coupled to an external environment characterized
by a continuum of states $|\omega \rangle$ with frequency $\omega$. The Hamiltonian of the full system is described by the multilevel Fano-Anderson (or Friedrichs-Lee) Hamiltonian \cite{r52tris}
\begin{equation}
\hat{H}=\sum_n \omega_n | a_n \rangle \langle a_n | + \int d \omega \omega | \omega \rangle \langle \omega| + f(t) \sum_n \left\{ g_n(\omega) | a_n \rangle \langle \omega| + g^*_n (\omega) | \omega \rangle \langle a_n| \right\}
\end{equation}
where $g_n(\omega)$ is the spectral amplitude that describes the coupling of the discrete state $|a_n \rangle$ with the continuum $| \omega \rangle$ and $f(t)$ defines the coupling strength, which is generally assumed to be time dependent. The Hamiltonian $\hat{H}$ is Hermitian provided that $f(t)$ is real. In this case the dynamics is unitary and the norm (probability) is conserved. However, rather generally we allow $f(t)$ to become complex, corresponding to a {\em non-Hermitian} discrete-continuum coupling and a non-unitary dynamics. It should be noted that a non-Hermitian coupling requires an hidden reservoir into which the discrete-continuum states can transfer particles/energy excitation, thus explaining the non-unitary dynamics and breakdown of population conservation. Non-Hermitian coupling by means of a complex (imaginary) function $f(t)$ is introduced here at a phenomenological level, i.e. without a detailed description of the 'hidden' reservoir which is eliminated from the dynamics and  that makes sense of an effective non-Hermitian discrete-continuum interaction. Indeed, the problem of time reversal of discrete states discussed in the following and based on the weak-coupling approximation does not require a detailed microscopic description of the non-Hermitian interaction. For example, photonic transport in coupled waveguides or optical resonators is often described in the framework of the Fano-Anderson model \cite{r54,r55}. Here the hidden reservoir is provided by the pumped atomic medium that provides instantaneous spatial optical gain and loss regions in the system and that makes the dynamics effectively non-Hermitian. In particular, an imaginary (non-Hermitian) coupling of wave guides can be implemented using the methods discussed in Refs. \cite{r48,r52,referee3}. Similarly, in non-Hermitian models of mesoscopic quantum transport imaginary potentials that act as source and sink for carriers  are phenomenologically introduced, without the need for a full description of the entire reservoir dynamics \cite{referee4,referee5}.
\\
If the state vector of the system $|\psi(t) \rangle$ is expanded on the discrete-continuum basis as
\begin{equation}
| \psi(t) \rangle = \sum_n c_n(t) |a_n \rangle + \int d \omega c(\omega,t) | \omega \rangle
\end{equation}   
 from the Schr\"odinger equation with $\hbar=1$ one obtains the following evolution equations for the amplitudes $c_n(t)$ and $c(\omega,t)$
 \begin{eqnarray}
 i \frac{dc_n}{dt} & = & \omega_n c_n(t)+f(t) \int d \omega g_n(\omega) c(\omega,t) \\
 i \frac{dc}{dt} & = & \omega c(\omega,t)+f(t) \sum_n g^*_n(\omega) c_n(t).
 \end{eqnarray}
Let us assume that at initial time the continuum of states is empty, i.e. $c(\omega,0)=0$. The formal integration of Eq.(4) yields
\begin{equation}
c(\omega,t)=-i \sum_n g^*_n(\omega) \int_0^t d\xi f(\xi) c_n(\xi) \exp[-i\omega(t-\xi)]
\end{equation}
After setting $c_n(t)=a_n(t) \exp(-i \omega_n t)$, substitution of Eq.(5) into Eq.(3) yields the following set of integro-differential equations for the amplitudes $a_n(t)$
\begin{equation}
 \frac{da_n}{dt}=-f(t) \sum_m \int_0^t d \xi f(\xi) a_m(\xi) \Phi_{n,m}(t-\xi) \exp(i \omega_nt-i \omega_m \xi)
\end{equation}
where we have introduced the memory functions
\begin{equation}
\Phi_{n,m}(\tau) \equiv  \int d \omega g_n(\omega) g_m^*(\omega) \exp(-i \omega \tau).
\end{equation}
The characteristic decay time $\tau_{men}$ of $\Phi_{n,m}(\tau)$ defines the memory time of the response function. Following a standard procedure, in the Weisskopf-Wigner (markovian) approximation, corresponding to a weak coupling $g_n(\omega) \rightarrow 0$, a broad and unstructured continuum, and a small change of $f(t)$ over the memory time $\tau_{mem}$, the integro-differential equations (6) can be replaced by the following set of differential equations
\begin{equation}
 \frac{da_n}{dt}=- f^2(t) \sum_m \Delta_{n,m} a_m(t) \exp [i(\omega_n-\omega_m)t]
\end{equation}
where we have set
\begin{equation}
\Delta_{n,m}  =   \int_0^{\infty} d \tau \Phi_{n,m}(\tau) \exp(i \omega_m \tau) = \pi g_n(\omega_m) g^*_m(\omega_m) - i P \int d \omega \frac{g_n(\omega) g_m^*(\omega) }{\omega-\omega_m}  \nonumber
\end{equation}
Equations (8) provide an effective non-Hermitian description of the dynamics in the subspace of the $N$ discrete states $|a_n \rangle$, in which the degrees of freedom of the continuum of states $| \omega \rangle$ have been eliminated in the markovian approximation. For an Hermitian and time-independent Hamiltonian $\hat{H}$, i.e. for $f(t)=1$,  the real and imaginary parts of the eigenvalues $\lambda_n$ of the effective coupling matrix $\mathcal{D}= \left\{ \Delta_{n,m} \right\}$ determine the decay rates and frequency shifts of the resonance modes. Such an effective non-Hermitian description is able to capture some important effects such as the existence of bound (trapping) states inside the continuum and Fano resonances arising from the interference of the discrete states via the continuum. It should be noticed that the above analysis is based on the rather standard markovian approximation, i.e. the solutions are obtained by assuming a weak coupling between the discrete states and  the continuum. This means that the validity of the results discussed in the next sections, i.e. time reversal and frozen dynamics by non-Hermitian flip, is strictly speaking restricted to the weak coupling regime. A more general approach, based on the Feshbach projection technique, would be needed to properly account for strong coupling and nonlinear processes involved in the non-Hermitian formalism \cite{r10,referee1,referee6}. However, for the sake of simplicity in the following analysis we will focus our attention to the simplest case of weak coupling.

\subsection{Time reversal and frozen dynamics by non-Hermitian flipping} 

Let us consider the evolution of the the amplitudes $a_n$ of the discrete states in the reduced subspace  $\{ |a_n \rangle \}$ as described by the effective non-Hermitian system (8). For the time interval $0 \leq t \leq T$, we assume a stationary and Hermitian discrete-continuum coupling $f(t)=1$. Since all eigenvalues of the $N \times N$ coupling matrix $\mathcal{D}=\Delta_{n,m}$ have positive or vanishing real part, a complete or fractional decay of amplitudes $a_n(t)$ for a long interaction time $T$ is observed. Fractional decay is found when one (or more) eigenvalues $\lambda_n$ of the matrix $\mathcal{D}$ have a vanishing  real part,  corresponding to the existence of one (or more) bound states arising from the destructive interference of different decay channels into the common continuum. From Eq.(8), it readily follows that time-reversed evolution can be achieved by flipping the discrete-continuum interaction from Hermitian to non-Hermitian, i.e. by assuming $f(t)=i$ in the interval $ T<t \leq 2T$, provided that the discrete states are frequency-degenerate, i.e. $\omega_n$ is independent of $n$ and equal for all discrete states $|a_n \rangle$. In this case, the complete or fractional decay dynamics of amplitudes $a_n$ is exactly reversed and at the final time $t=2T$ the discrete states return to their initial excitation, i.e. a Loschmidt echo is realized. However, since in the interval $T<t<2T$ the Hamiltonian $\hat{H}$ is not Hermitian, the continuum of states at final time $t=2T$ does not generally return to its initial empty excitation, rather some excitation is left in the continuum as a signature of non-unitary dynamics. It should be noted that exact time reversal strictly requires frequency degeneracy of the discrete states. However, approximate time reversal can be observed more generally for non-degenerate levels as well, provided that $|\omega_n-\omega_m|$ is much larger than the absolute value of any eigenvalue $\lambda_n$ of the matrix $\mathcal{D}$, as it readily follows from application of the rotating-wave approximation to Eq.(8).\par
 Interestingly, for arbitrary frequencies $\omega_n$ of discrete states a frozen dynamics can be observed by periodically switching the discrete-continuum interaction $f(t)$ from Hermitian ($f=1$) to non-Hermitian ($f=i$) at time intervals $T$, provided that $T$ is smaller than the characteristic evolution time of the system (8), i.e. provided that $T|\lambda_n| \ll 1$ and $T|\omega_n-\omega_m| \ll 1$ for any $n,m=1,2,...,N$. Such a kind of frozen dynamics by periodic time reversal of the dynamics via non-Hermitian flipping of the interaction bears some analogy with decay suppression in quantum Zeno dynamics based on frequent observations or dynamical decoupling methods, where transitions between the discrete states and the continuum are effectively suppressed. However, our scheme holds in the Weisskopf-Wigner (markovian) limit and the dynamics is frozen for the discrete states $|a_n \rangle$ solely, whereas excitation in the continuum of states shows a secular growth as a signature of non-Hermitian dynamics. Such a secular growth of excitation in the continuum results from energy/population feeding from the hidden reservoir and will be exemplified in the next section. It is also worth commenting that in the ordinary Hermitian discrete-continuum interaction the weak-coupling approximation means that the decay process of the discrete states occurs without memory effects, i.e. it is markovian, and that loss of memory implies an irreversible decay into the continuum, without the possibility of time reversing the dynamics neither of observing echoes effects. Therefore, time reversal dynamics of the discrete states obtained after non-Hermitian switching of the discrete-continuum coupling can be regarded as a sort of memory effect. However, we stress that in the weak-coupling limit time-reversal of the discrete states does not require the knowledge of the past dynamics, since the entire dynamics before and after non-Hermitian flip is instantaneous and described by differential (rather than integro-differential) equations. Neither memory effects in the hidden reservoir are required to explain time reversal. In fact, the collective decay dynamics of the discrete states in the early (Hermitian) stage is reversed because the coupling matrix $\{ \mathcal{D}\}$ is exactly replaced by its time reversal $\{ - \mathcal{D}\}$ after non-Hermitian flip ($f^2=-1$), so as the collective decay is followed by a collective amplification until the initial state is retrieved. In other words, time reversal does not arise from refocusing of the excitation from the continuum of states $| \omega \rangle$ into the discrete states $|a_n \rangle$, like in ordinary  Loschmidt echo which would require to keep memory of the entire dynamics. Re-excitation of the discrete states to their original condition after non-Hermitian flip of the coupling is assisted by extra energy provided by the hidden reservoir, however such a time reversal does not require to keep memory of the dynamics and the entire discrete-continuum system is not time reversed.

\section{Loschmidt echo and frozen dynamics of discrete states attached to a tight-binding continuum}
To exemplify the concept of time-reversal via non-Hermitian flipping, let us consider an idealized system describing the decay dynamics of $N$ discrete states side-coupled to a one-dimensional tight-binding lattice, as schematically shown in Fig.1. Such a system can describe, for example, mesoscopic quantum transport in a tight-binding quantum wire with attached $N$ quantum dots \cite{r53}, or photonic transport in $N$ waveguides or optical cavities side-coupled by evanescent tunneling to an array of waveguides or optical cavities \cite{r54,r55,r56}. Indicating by $c_n$ ($n=1,2,...,N$) and $b_{\alpha}$ ($\alpha=0, \pm 1, \pm 2, ...$) the amplitudes of excitation in the discrete $N$ states $|a_{n} \rangle$ and in the Wannier sites $|\alpha \rangle$ of the tight-binding lattice, respectively,  the evolution equations for the amplitudes $c_n$ and $b_{\alpha}$ read 
\begin{eqnarray}
i \frac{dc_n}{dt} & = & \omega_n c_n-f(t) \kappa_n b_{\alpha_n} \\
i \frac{db_{\alpha}}{dt} & = & -\kappa (b_{\alpha+1}+b_{\alpha-1})-f(t) \sum_{n=1}^{N} \kappa_n c_n \delta_{\alpha, \alpha_n}.
\end{eqnarray}
In the above equations, $\omega_n$ is the frequency offset of the discrete state $|a_n \rangle$ from the center of the tight-binding lattice band, $\kappa$ is the hopping amplitude between adjacent Wannier sites in the lattice, $\kappa_{n}$ is the hopping amplitude between the discrete state $|a_n \rangle$ and the Wannier site $| \alpha_n \rangle$ in the lattice, and $f(t)=1$ [$f(t)=i$] for Hermitian [non-Hermitian] coupling.  The frequencies of the discrete states are assumed to be embedded into the continuous spectrum of the tight-binding lattice band $ -2 \kappa \leq \omega \leq 2 \kappa$, i.e. $|\omega_n| < 2 \kappa$. The equivalence of dynamical equations (10-11) with the ones (3-4)  presented in the previous section can be readily established by the introduction of the Bloch basis (rather than the Wannier basis) for the states of the tight-binding continuum \cite{r54,r55}. In the Weisskopf-Wigner (markovian) approximation, the amplitudes $b_{\alpha} $ of the tight-binding lattice can be eliminated from the dynamics following the general procedure outlined in the previous section.  After setting $c_n(t)=a_n(t) \exp(-i \omega_n t)$, the evolution equations for the amplitudes $a_n$ are thus found to be described by the effective non-Hermitian system (8) with the elements of the coupling matrix $\mathcal{D}$ given by Eq.(9). Analytical computation of the integrals yields
\begin{equation}
\Delta_{n,m}= \kappa_n \kappa_m i^{|\alpha_n-\alpha_m|} \frac{\left[ \left( 4 \kappa^2-\omega_m^2 \right)^{1/2} +i \omega_m \right]^{|\alpha_n-\alpha_m|}}{(2 \kappa)^{|\alpha_n-\alpha_m|} \left( 4 \kappa^2 -\omega_{m}^{2} \right)^{1/2}}.
\end{equation}
As an example, Fig.2 shows time-reversal in the subsystem of $N=3$ discrete states via non-Hermitian flipping as obtained by numerical simulations of Eqs.(10-11) for the three sites $|a_1 \rangle$, $|a_2 \rangle$ and $| a_3 \rangle$ side-coupled to the tight-binding lattice at the sites $|-1 \rangle$, $|0 \rangle$ and $|1 \rangle$ [see the inset of Fig.2(b)]. Parameter values used in the simulations are $\kappa_1 / \kappa=0.0375$, $\kappa_2/ \kappa=0.025$, $\kappa_3 / \kappa=0.05$, $\omega_1 / \kappa=\omega_2 / \kappa =\omega_3 / \kappa=0$, and $\kappa T=200$. The initial excitation of the system is $a_1(0)=1/ \sqrt{3}$, $a_2(0)=-i/ \sqrt{3}$, $a_3(0)=-1/\sqrt{3}$, $b_{\alpha}(0)=0$, corresponding to an empty continuum. The solid curves in Fig.2(a) depict the temporal evolution of the populations $P_l(t)=|a_l(t)|^2$ ($l=1,2,3$) of the three discrete states, whereas the curves in Fig.2(b) show the evolution of the population in the continuum $P_{c}(t)=\sum_{\alpha=-\infty}^{\infty}  |b_{\alpha}(t)|^2$ and of the total population $P_{tot}(t)=P_1(t)+P_2(t)+P_3(t)+P_{c}(t)$. The dashed curves in Fig.2(a), almost overlapped with the solid ones, show the behavior of the populations of the three discrete states as predicted by the reduced non-Hermitian system (8) in the markovian approximation, with the elements $\Delta_{n,m}$ of the effective coupling matrix $\mathcal{D}$ computed using to Eq.(12). Note the excellent accuracy of the reduced model to predict the full numerical results, where the continuum degrees of freedom are not eliminated. Figure 2(a) clearly shows that the initial decay of populations in the initial stage of the dynamics, i.e. for the time interval $0<\kappa t< 200$ with Hermitian coupling $f(t)=1$, is time reversed in the second stage of the dynamics, i.e. in the time interval $200<\kappa t< 400$ after non-Hermitian flipping of the coupling ($f(t)=i$). Hence a Loschmidt echo is observed for the discrete states. However, Fig.2(b) indicates that the population in the continuum shows a secular growth without returning to its initial vanishing value. Such a behavior is a clear signature of the non-unitary evolution of the system after non-Hermitian flipping of the coupling. The accuracy of time reversal for the subsystem of discrete states solely can be quantified by means of the fidelity $\mathcal{F}$\begin{equation}
\mathcal{F}(t)=\frac{\left| \sum_{n=1}^{N} a_n(0)a_n^*(t) \right|}{ \sum_{n=1}^{N} |a_n(0)|^2}
\end{equation}
which is shown in Fig.2(c). Perfect time reversal corresponds to $\mathcal{F}=1$, i.e. $a_n(t)=a_n(0)$, whereas deviations of $\mathcal{F}$ from one is a signature of imperfect time reversal. Note that, since the dynamics is not unitary, the fidelity could even become larger than one, indicating that population in the subsystem of discrete states can become larger than the initial one.\\ 
According to the theoretical analysis outlined in the previous section, time reversal can be obtained provided that the discrete states are energy degenerate (as in Fig.2), or approximately when non degeneracies of the discrete states correspond to frequency detunings much larger than the effective couplings among them. An example of time reversal obtained for non-degenerate states is shown in Fig.3(a). On the other hand, when the frequency mismatches of discrete states are comparable than the effective state couplings (which are determined by the order of magnitude of the elements of the coupling matrix $\mathcal{D}$) non-Hermitian flip fails to realize time reversal of the discrete states, as shown in Figs.3(b) and (c). Finally, Fig.4 shows an example of frozen dynamics of the subsystem of the $N=3$ discrete states as obtained by periodic alternation of Hermitian ($f=1$) and non-Hermitian ($f=i$) couplings at times $t=T,2T,3T,...$ for the same parameter values of Fig.3(b) and for $ \kappa T=8$.

   \section{Conclusions}
Controlling the dynamical properties of $N$ discrete states coupled to a continuum is of major interest in different areas of quantum and classical physics. Elimination of the continuum degrees of freedom in the Weisskopf-Wigner approximation leads to an effective non-Hermitian Hamiltonian description for the finite-dimensional subspace of discrete states, which is capable of predicting important interference effects such as Fano resonances and the existence of bound states in the continuum. In principle, time reversal of the dynamics for both the discrete and the continuum of states requires inverting the sign of the Hamiltonian.  Here we suggested a different route toward time reversal of the subsystem of the discrete states coupled to a common continuum, based on flipping of the discrete-continuum coupling from an Hermitian to a non-Hermitian interaction \cite{r52}, thus resulting in a non unitary dynamics. While simply changing the sign of the full Hamiltonian yields time reversal of both the discrete and the continuum of states, non-Hermitian flip of the interaction reverses the temporal evolution of the discrete states while the continuum of states is not reversed. Exact time reversal requires frequency degeneracy of the discrete states, or large frequency mismatch among the discrete states as compared to the strength of indirect coupling mediated by the continuum. Interestingly, periodic and frequent switch of the discrete-continuum coupling from Hermitian to non-Hermitian interaction results in a frozen dynamics of the subsystem of discrete states. The concepts of time reversal and frozen dynamics in a subsystem of discrete states based on non-Hermitian flip have been exemplified by considering the decay dynamics of $N$ discrete states side-coupled to a linear tight-binding lattice. In such a system, which could describe photonic transport in coupled optical resonators or waveguides \cite{r54,r55,r56}, an imaginary (non-Hermitian) coupling of the discrete states with the lattice can be realized by non-Hermitian engineering methods, as discussed in previous works \cite{r48,r52},
The present results shed new light into time reversal properties of non-Hermitian systems, revealing how non-Hermitian flips can be exploited to realize time reversal of a subsystem of discrete states as well as to freeze the dynamics.\par
{\em Conflict of interest}: The author declares that he has no conflict of interest.

 \begin{figure}[b]
\includegraphics[width=10cm]{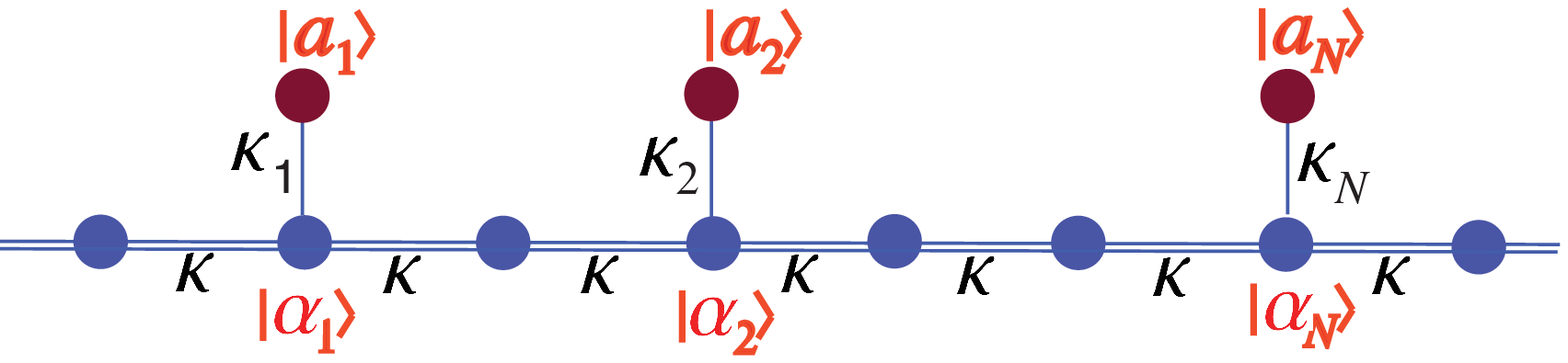}
\caption{(Color online). Schematic of a linear tight-binding lattice with $N$ localized states $|a_1 \rangle$, $|a_2 \rangle$, ... $|a_N \rangle$ attached at the linear lattice at the sites $|\alpha_1 \rangle$, $|\alpha_2 \rangle$, ..., $|\alpha_N \rangle$ with hopping rates $\kappa_1$, $\kappa_2$, ..., $\kappa_N$. The tight-binding lattice shows a band of delocalized modes (Bloch waves) in the frequency range $-2 \kappa \leq \omega \leq 2 \kappa$, where $\kappa$ is the hopping amplitude between adjacent sites in the lattice. The frequency offsets of the localized states from the center of the tight-binding lattice band are $\omega_1$, $\omega_2$, ... , $\omega_N$.} 
\end{figure}
\begin{figure}[b]
\includegraphics[width=16cm]{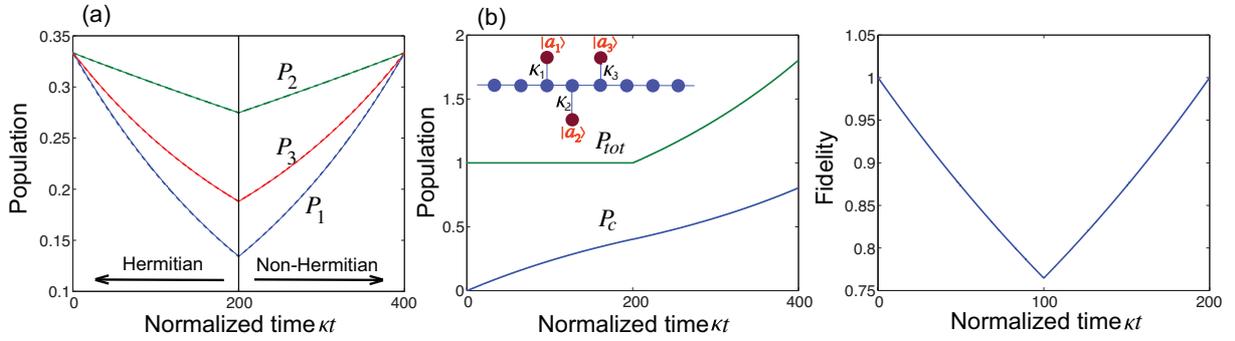}
\caption{(Color online). Example of time reversal of $N=3$ frequency-degenerate discrete states side-coupled to a tight-binding continuum after non-Hermitian flipping of the coupling. (a) Numerically-computed temporal evolution of the populations $P_1(t)$, $P_2(t)$ and $P_3(t)$ in the three discrete states. Parameter values are given in the text. Flipping of the coupling is achieved at time $\kappa t=200$. The dashed curves in the figure, almost overlapped with the solid ones, show the evolution of the populations as predicted in the markovian approximation [Eqs.(8) and (12)]. (b) Evolution of the population in the continuum ($P_c$) and of the total population ($P_{tot}$). Note that the dynamics is unitary, corresponding to population conservation, for $\kappa t < 200$, whereas a secular growth of population in the continuum is observed after non-Hermitian flip of the coupling ($\kappa t > 200$), which is a signature of non-unitary dynamics. The inset in (b) shows schematically the three discrete states attached to the tight-binding lattice. (c) Temporal evolution of the fidelity $\mathcal{F}(t)$.} 
\end{figure}
\begin{figure}[b]
\includegraphics[width=16cm]{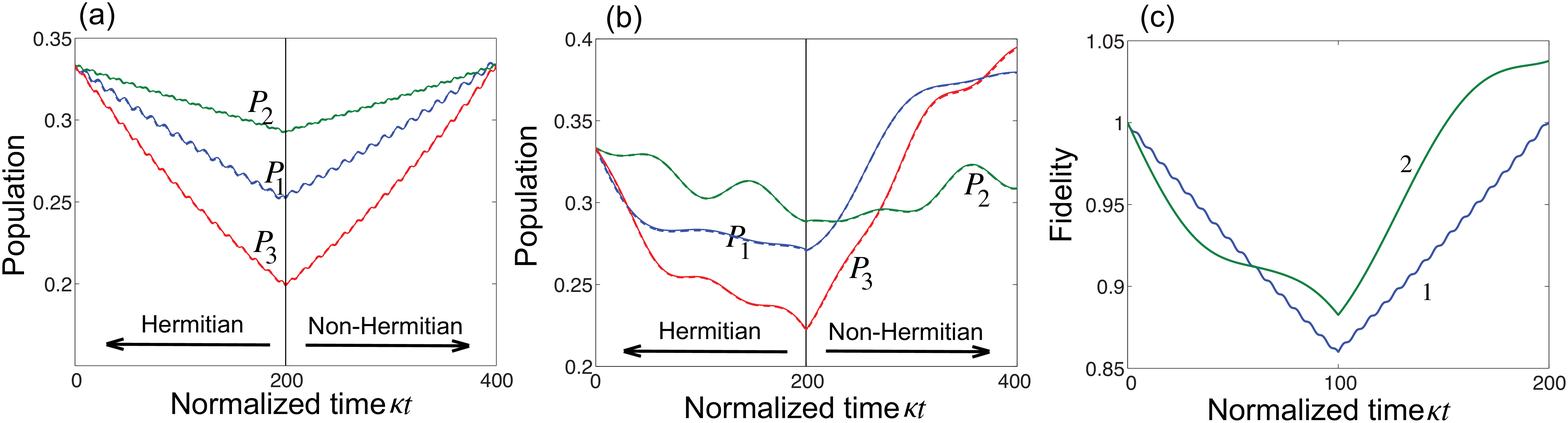}
\caption{(Color online). (a) Same as Fig.2(a), but for $\omega_1 / \kappa=0$, $\omega_2 / \kappa=0.5$ and $\omega_3 / \kappa=-0.5$. (b) Same as Fig.2(a), but for 
 $\omega_1 / \kappa=0$, $\omega_2 / \kappa=0.05$ and $\omega_3 / \kappa=-0.025$. (c) Temporal evolution of the fidelity $\mathcal{F}(t)$. Curve 1 referes to (a), whereas curve (b) refers to (b). The circumstance that in (b) the fidelity increases above one [see curve 2 in (c)] means that the population in the subsystem of discrete states becomes larger than the initial one.} 
\end{figure}
\begin{figure}[b]
\includegraphics[width=14cm]{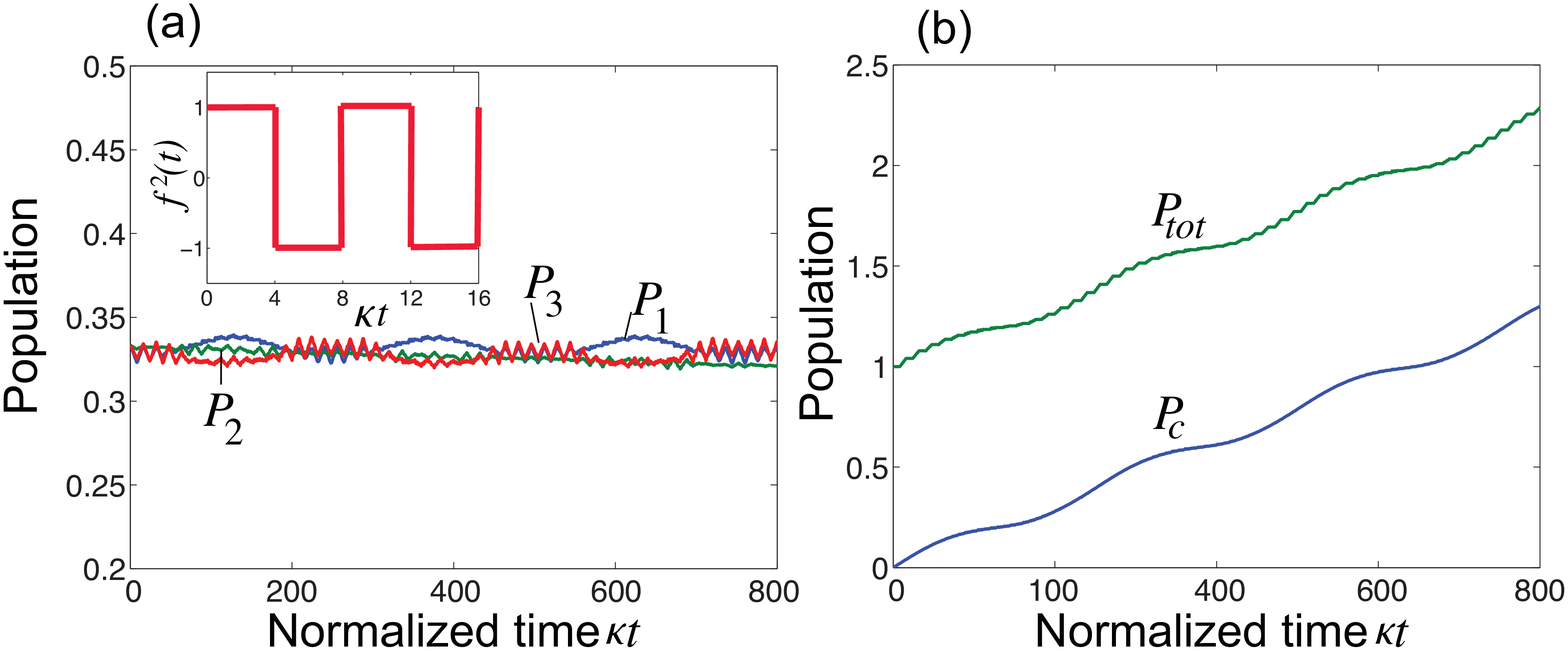}
\caption{(Color online). Frozen dynamics in the subsystem of the three discrete states [inset of Fig.2(b)] as obtained by periodic alternation of the coupling from Hermitian ($f^2=1$) to non-Hermitian ($f^2=-1$) interaction.  (a) Numerically-computed evolution of the populations $P_1(t)$, $P_2(t)$ and $P_3(t)$ of the three discrete states for 
 $\omega_1 / \kappa=0$, $\omega_2 / \kappa=0.05$ and $\omega_3 / \kappa=-0.025$ and for periodic alternation of $f^2$ at times $t=T,2T,3T,...$ with $\kappa T=4$ (see the inset). The couplings $\kappa_1$, $\kappa_2$ and $\kappa_3$ are as in Fig.2. (b) Evolution of the population in the continuum ($P_c$) and of the total population ($P_{tot}$). Note the secular growth of the population in the continuum, which is a clear signature of non-unitary dynamics. } 
\end{figure}


\begin{thebibliography}{00}


\bibitem{r1}
 Gamow G. Zur Quantentheorie des Atomkernes. Z Physik 1928; 51:204-212.
 
 \bibitem{r2}
 Weisskopf V, Wigner E. Berechnung der natuKrlichen Linienbreiten auf Grund der Diracschen Lichttheorie. Z
Phys 1930; 63: 54-73.
 
 \bibitem{r3}
 Feshbach H. Unified Theory of Nuclear Reactions. Ann Phys 1958; 6: 357-399.
 
 \bibitem{r4}
Sternheimt MM, Walker JF.  Non-Hermitian Hamiltonians, Decaying States, and Perturbation Theory.
 Phys Rev C 1972; 6:114-121.
  
  \bibitem{r5}
  Cohen-Tannoudji C, Dupont-Roc J, Grynberg G. Atom-photon interactions: basic processes and applications. New York: John Wiley \& Sons Inc; 1992.
  
 \bibitem{r6}
 Dittes FM. The decay of quantum systems with a small number of open channels. Phys Rep 2000; 339:215-316.
 
 \bibitem{r7}
 Breuer HP, Petruccione F. The Theory of Open Quantum Systems. New York: Oxford University Press;  2002.
 
\bibitem{r8}
Moiseyev N. Non-Hermitian Quantum Mechanics. Cambridge: Cambridge University Press; 2011.

\bibitem{r9}
Rotter I. A non-Hermitian Hamilton operator and the physics of open quantum systems. J Phys A 2009; 42: 153001.

\bibitem{r10}
Rotter I, Bird JP. A review of progress in the physics of open quantum systems: theory and experiment. Rep Prog Phys 2015; 78: 114001.

\bibitem{referee1}
Eleuch H, Rotter I. Resonances in open quantum systems. Phys Rev A 2017; 95: 022117.

\bibitem{referee2}
Yoon Y, Kang M-G, Morimoto T, Kida M, Aoki A, Reno J L, Ochiai Y, Mourokh L, Fransson J, Bird J P.
Coupling Quantum States through a Continuum: A Mesoscopic Multistate Fano Resonance. Phys. Rev. X 2012; 2: 021003.

\bibitem{r11}
 Bagarello F, Passante R, Trapani C. Non-Hermitian Hamiltonians in Quantum Physics. Berlin: Springer; 2016.

\bibitem{r12}
Wunner G, Pelster A. Self-organization in Complex Systems: The Past, Present, and Future of Synergetics.
Berlin: Springer; 2012.

\bibitem{r13}
 Gardas B, Deffner S, Saxena A. Non-Hermitian quantum thermodynamics. Sci Rep 2016; 6: 23408.
        
\bibitem{r14}    
Nesterov AI. Berman G P. Quantum search using non-Hermitian adiabatic evolution. Phys Rev A 2012; 86: 052316.

\bibitem{r15}
Croke S. PT-symmetric Hamiltonians and their application in quantum information. Phys. Rev. A 2015; 91: 052113.

\bibitem{r16}
Datta S. Electronic Transport in Mesoscopic System. Cambridge: Cambridge University Press; 1995.

\bibitem{r17}
Sadreev AF,  Rotter I. S-matrix theory for transmission through billiards in tight-binding approach . J. Phys. A 2003; 36: 11413.

\bibitem{r18}
Sadreev AF, Bulgakov EN, Rotter I. Trapping of an electron in the transmission through two quantum dots coupled by a wire. JETP Lett 2005; 82: 498-503.
     
\bibitem{r19}   
Rotter I, Sadreev AF. Avoided level crossings, diabolic points, and branch points in the complex plane in an open double quantum dot. Phys Rev E 2005; 71: 036227.
     
\bibitem{r20}     
Celardo GL, Kaplan L. Superradiance transition in one-dimensional nanostructures: An effective non-Hermitian Hamiltonian formalism.
Phys Rev B 2009; 79: 155108.

\bibitem{r21}
Giusteri GG,  Mattiotti F, Celardo GL. Non-Hermitian Hamiltonian approach to quantum transport in disordered networks with sinks: validity and effectiveness.
Phys Rev B 2015: 91; 094301.

\bibitem{r22}
Ostahie B, Ni?a M, Aldea A. Non-Hermitian approach of edge states and quantum transport in a magnetic field. Phys Rev B 2016; 94: 195431.

\bibitem{r23}
Siegman A E.  Excess spontaneous emission in non-Hermitian optical systems. I. Laser amplifiers. Phys Rev A 1999; 39: 1253-1263.

\bibitem{r24}
Berry MV. Physics of Nonhermitian Degeneracies. Czech J Phys 2004; 54: 1039-1047.

\bibitem{r25}
Alaeian H, Dionne JA. Non-Hermitian nanophotonic and plasmonic waveguides. Phys Rev B 2014; 89: 075136.

\bibitem{r26}
Longhi S, Gatti D, Della Valle G. Robust light transport in non-Hermitian photonic lattices. Sci Rep 2015; 5: 13376.

\bibitem{r27}
Savoia S, Castaldi G, Galdi V. Complex-coordinate non-Hermitian transformation optics. J Opt 2016; 18: 044027.

\bibitem{r28}
Amir A, Hatano N, Nelson DR. Non-Hermitian Localization in Biological Networks. Phys Rev E 2016; 93: 042310.

\bibitem{r29}
Nesterov AI, Berman GP, Bishop AR. Non-Hermitian approach for modeling of noise-assisted quantum electron transfer in photosynthetic complexes.
Fortschr Phys. 2012; 61:1-16.

\bibitem{r30}
 Ferrari D, Celardo GL, Berman GP et al. Quantum Biological Switch Based on Superradiance Transitions.
J Phys Chem C 2014; 118: 20-26.

\bibitem{r31}
Bender CM, Boettcher S. Real Spectra in Non-Hermitian Hamiltonians Having PT Symmetry. Phys Rev Lett 1998; 80: 5243-5246.

\bibitem{r32}
Bender CM, Brody DC, Jones HF. Complex Extension of Quantum Mechanics. Phys Rev Lett 2002; 89: 270401.

\bibitem{r33}
Bender CM, Making sense of non-Hermitian Hamiltonians. Rep Prog Phys 2007; 70: 947-1018. 

\bibitem{r34}
Mostafazadeh A. Pseudo-Hermiticity versus PT symmetry: The necessary condition for the reality of the spectrum of a non-Hermitian Hamiltonian. J Math Phys 2002; 43: 205-214.

\bibitem{r35}
Mostafazadeh A. Pseudo-Hermiticity versus PT symmetry. II. A complete characterization of non-Hermitian Hamiltonians with a real spectrum. J Math Phys 2002; 43: 2814-2816.

\bibitem{r36}
Mostafazadeh A. Non-Hermitian Hamiltonians with a Real Spectrum and Their Physical Applications. Pramana J Phys 2009; 73: 269-277.

\bibitem{r37}
Mostafazadeh A. Pseudo-Hermitian representation of quantum mechanics. Int. J. Geom Methods Mod Phys (2010); 7: 1191-1306. 

\bibitem{r38}
Gorin T, Prosen T, Seligman TH et al.  Dynamics of Loschmidt echoes and fidelity decay. Phys. Rep. 2006; 435: 33-156 

\bibitem{r39}
Goussev A, Jalabert RA, Pastawski HM et al. Loschmidt echo. Scholarpedia 2012; 7: 11687.

\bibitem{r40}
Goussev A, Jalabert RA, Pastawski HM et al. Loschmidt echo and time reversal in complex systems. Phil Trans R Soc A 2016; 374: 20150383.

\bibitem{r41}
Hahn EL. Spin Echoes. Phys Rev 1950; 80 580-594.

\bibitem{r42}
Martin J, Georgeot B, Shepelyansky DL. Cooling by Time Reversal of Atomic MatterWaves. Phys Rev Lett 2008; 100: 044106.

\bibitem{r43}
Ullah A and Hoogerland MD. Experimental observation of Loschmidt time reversal of a quantum chaotic system. Phys Rev E 2011; 83: 046218.

\bibitem{r44}
Fink M, Cassereau D, Derode A et al. Time-reversed acoustics. Rep Prog Phys 2000; 63: 1933-1995.

\bibitem{r45}
Lerosey G, de Rosny J, Tourin A et al. Time Reversal of ElectromagneticWaves. Phys Rev Lett 2004; 92: 193904.

\bibitem{r46}
Bacot V, Labousse M,  Eddi A et al. Time reversal and holography with spacetime transformations. Nat Phys 2016; 12: 972-977.

\bibitem{r47}
Bender CM, Brandt SF, Chen JH et al. Ghost busting: PT-symmetric interpretation of the Lee model. Phys Rev D 2005; 71: 025014.
D 71, 025014 (2005).

\bibitem{r48}
Longhi S.  Invisibility in non-Hermitian tight-binding lattices. Phys Rev A 2010; 82: 032111.

\bibitem{r49}
Zhang XZ, Jin L, Song Z. Self-sustained emission in semi-infinite non-Hermitian systems at the exceptional point. Phys Rev A 2013; 87: 042118.

\bibitem{r50}
Longhi S. Bound states in the continuum in PT-symmetric optical lattices. Opt Lett 2014; 39: 1697-1700.

\bibitem{r51}
Bagarello F, Lattuca M, Passante R et al. Non-Hermitian Hamiltonian for a Modulated Jaynes-Cummings Model with PT Symmetry. Phys Rev A 2015; 91: 042134.

\bibitem{r52}
Longhi S. Non-Hermitian tight-binding network engineering. Phys Rev A 2016; 93: 022102.

\bibitem{r52bis}
Longhi S. Suppression of quantum decay via non-Hermitian coupling. 2016; arXiv:1701.00483.

\bibitem{r52tris}
Miyamoto M. Bound-state eigenenergy outside and inside the continuum for unstable multilevel systems. Phys Rev A 2005; 72:063405.	

\bibitem{r54}
Dreisow F, Szameit A, Heinrich M et al. Adiabatic transfer of light via a continuum in optical waveguides. Opt Lett 2009; 34: 2405-2407.

\bibitem{r55}
Dreisow F, Szameit A, Heinrich M et al. Decay Control via Discrete-to-Continuum Coupling Modulation in an Optical Waveguide System. Phys Rev Lett 2008; 101: 143602.

\bibitem{referee3}
Alexeeva N V, Barashenkov I V, Rayanov K, Flach S. Actively coupled optical waveguides. Phys. Rev. A 2014; 89: 013848.

\bibitem{referee4}
 Berggren K-F, Yakimenko I I, Hakanen. Modeling of open quantum dots and wave billiards using imaginary potentials for the source and the sink. J. New J. Phys. 2010; 12. 073005.
 
 \bibitem{referee5}
 {Elenewski J E, Chen H. Real-time transport in open quantum systems from PT-symmetric quantum mechanics. Phys. Rev. B 2014; 90: 085104.}
 
 \bibitem{referee6}
 Bulgakov E N, Rotter I, Sadreev A F. Comment on Bound-state eigenenergy outside and inside the continuum for unstable multilevel systems. Phys. Rev. A 2007; 77:067401.

\bibitem{r53}
Orellana P, Dominguez-Adame F. Conductance control in quantum wires by attached quantum dots. Phys Stat Sol A 2006; 203: 1178-1181.

\bibitem{r56}
 Hafezi M,	Demler EA, Lukin MD et al. Robust optical delay lines with topological protection. Nat Phys 2001; 7: 907?912.
 


\end{thebibliography}
\end{document}